\documentstyle[12pt]{article}                                 

\begin{document}

\newcommand{\bea}{\begin{eqnarray}}
\newcommand{\eea}{\end{eqnarray}}
\newcommand{\be}{\begin{equation}}
\newcommand{\ee}{\end{equation}}
\newcommand{\ads}[1]{{\rm AdS}_{#1}}

\newcommand{\yv}{\vec{y}}

\title{
\begin{flushright}
\begin{small}
hep-th/9811120\\
EFI-98-56 , FERMILAB-Pub-98/358-T \\
November 1998 \\
\end{small}
\end{flushright}
\vspace{1.cm}
The Coulomb Branch of Gauge Theory from Rotating Branes \\
}
\author{Per Kraus and Finn Larsen
\thanks{pkraus@theory.uchicago.edu, flarsen@theory.uchicago.edu}\\
\small the Enrico Fermi Institute\\
\small 5640 S. Ellis Ave.\\
\small Chicago, IL 60637\\
\and 
Sandip P. Trivedi
\thanks{trivedi@fnth23.fnal.gov}\\
\small Fermi National Accelerator Laboratory \\
\small P.O.Box 500 \\
\small Batavia, IL 60510\\
}

\date{}
\maketitle

\begin{abstract}
At zero temperature the Coulomb Branch of ${\cal N}=4$ super Yang-Mills theory
is described in supergravity by multi-center solutions with
D3-brane charge. At finite temperature and chemical potential
the vacuum degeneracy
 is lifted, and minima of the free energy are shown to have a
supergravity description as rotating black D3-branes.
 In the extreme limit these solutions single out
preferred points on the moduli space that can be interpreted as
simple distributions of branes --- for instance, a uniformly charged
planar disc. We exploit this geometrical representation to study the 
thermodynamics of rotating black D3-branes.  
The low energy excitations of the system appear to be governed 
by an effective string theory which is related to the singularity
in spacetime. 
\end{abstract}

\section{Introduction}
\label{sec:intro}

It has become evident that many questions concerning large $N$ gauge
theories can be answered via supergravity~\cite{juanads}.  
Among the many applications
of this approach, one can compute correlation functions of the gauge 
theory~\cite{polyakovads,wittenads},
map out its phase structure and study general thermodynamic 
properties~\cite{imsy,wittenads2,barbonth,mart1,mart2}.
In almost all investigations conducted so far the gauge theories have been
studied at the origin of moduli space, where there is  an enhanced
superconformal symmetry.  At this special point the group theory of the
superconformal algebra can be brought to bear, leading to many exact
results even at strong coupling.  Such field theories are described
by gravity in anti-de Sitter space (AdS).

However, it is also desirable to extend the investigations away from the
origin of moduli space.  Giving expectation values to certain scalar 
fields moves one onto the Coulomb branch --- a space of maximally
supersymmetric but nonconformal vacua.  We would like to know how the physics
of the Coulomb branch can be addressed using supergravity.  As a complementary
motivation, for the purposes of understanding quantum gravity via 
gauge theory, one would like to have access to geometries besides just AdS.
While linearized perturbations around AdS are well understood in both
the gravity and gauge theory contexts, less has been said about large 
departures from AdS which involve the nonlinear aspects of gravity in an 
important way.

In this work we study the Coulomb branch of ${\cal N}=4$ super Yang-Mills
theory.  The origin of moduli space represents a collection of coincident
D3-branes, whose near horizon geometry is $AdS_5 \times S^5$.  Separating
some of the branes in the transverse space moves the gauge theory onto the
Coulomb branch.  One would like to associate each point on the Coulomb 
branch with some new supergravity solution which is {\em asymptotically}
$AdS_5 \times S^5$.  Indeed, solutions representing multiple D3-branes
at distinct points are well known, and referred to as multi-center metrics.
Such supergravity solutions are written in terms of a harmonic function
which has sources at the positions of the D3-branes.  Thus, as we review
in Section~\ref{sec:coulomb}, 
there is a simple one-to-one correspondence between points on
the Coulomb branch and multi-center geometries (see also
~\cite{juanads,bulkbranes,bilal,wu,tseyttherm}).

Once the moduli space has been understood we consider gauge theory on the 
Coulomb branch at finite temperature.  Then supersymmetry is broken
and the vacuum degeneracy is removed; 
the gauge theory settles to points which minimize the free
energy of the system.  If temperature is the only
thermodynamic parameter present, then one expects that the free energy will 
be minimized at
the origin of moduli space:  away from the origin, certain fields acquire
masses and so contribute less to the entropy.  To investigate the theory
away from the origin we can include a chemical potential
for charge under the global $SO(6)$ R symmetry group of the ${\cal N}=4$ 
theory.  Physically, this means adding angular momentum in the space
transverse
to the collection of D3-branes.  Now we expect the free energy to be
minimized away from the origin, and we would like to know precisely where.   
At weak coupling
this can be answered through a perturbative computation of the free energy.
At strong coupling we turn to supergravity for the answer.

At zero temperature we noted that there was a supergravity solution 
corresponding to each point on the Coulomb branch.  The situation at
finite temperature is entirely different. From the 
no-hair theorems we expect that there is a unique solution corresponding to
each value of the mass, charge, and angular momentum.  The solution of
interest to us -- nonextremal, rotating D3-branes -- has been derived
up to duality in~\cite{cy96d}, and it is given in an Appendix. 
According to our discussion, 
this solution represents excited  states of the gauge theory at points
 in moduli space that minimize
the free energy.  

We identify these special points in moduli space in Section \ref{sec:sugra}
by taking the
extreme limit of the rotating D3-brane geometry~\cite{russorot}, 
since in this limit
the solution must reduce to some multi-center solution.  
The answer turns out to be very simple. It depends on how many of 
the three independent $SO(6)$ angular momentum parameters
$l_{1,2,3}$ are nonvanishing.  
$(l_1 \neq 0, l_{2,3}=0)$ corresponds to a uniform
distribution of D3-branes on a two dimensional disk of radius $l_1$; 
$(l_{1,2} \neq 0, l_{3}=0)$ corresponds to a distribution on a 
three dimensional ellipsoidal shell, 
$(y_1^2+y_2^2)/l_1^2 +(y_3^2+y_4^2)/l_2^2=1$. The
general case $l_{1,2,3} \neq 0$ is more subtle, as we discuss;
at this point we simply note that the distribution is contained
within a five dimensional ellipsoid.  For completeness, we also derive
the extreme limits of rotating M2 and M5 brane solutions;  the results
are qualitatively similar.

  From the nonextreme solutions one can deduce various thermodynamic 
quantities, and these become predictions for the large $N$ gauge
 theory on the Coulomb branch. We discuss these predictions in 
Section~\ref{sec:thermo}. In some 
respects the
features revealed by black hole thermodynamics  agree 
qualitatively with expectations for the gauge theory; in other respects,
they reveal some intriguing new aspects. 
For example, in the case where only one angular momentum parameter $l_1$
is non-zero, and the extremal configuration consists of branes distributed
on a disk,  we find that low-energy excitations above the 
extremal state
are governed by an effective string theory with a string tension that 
is determined by the Higgs VEV. 

The extreme solutions we discuss are typically singular on the surface 
containing the D3-branes. For a disk configuration we show that such
 singularities have a gauge theory 
interpretation: as the scale decreases along the renormalization 
group flow, the effective field theory changes from the ultraviolet 
${\cal N}=4$ theory to the infra-red effective string theory mentioned above. 
One expects the transition between the two to occur at a scale of order
the string tension in the effective string  theory. Using a Wilson loop  probe
we argue, in Section \ref{sec:sing},  
that this energy scale in the gauge theory 
corresponds to the radial 
position of the singularity in the bulk theory.

A particularly interesting multi-center solution is obtained by distributing 
the D3-branes uniformly on a five sphere of radius $l$. Then 
the supergravity solution is given by flat ten dimensional
spacetime inside the sphere, and $AdS_5 \times S^5$ outside.  In the
gauge theory, we expect that 
the inner region is represented by the low energy effective
theory valid below scale $l$.

This paper is organized as follows. In the following Section 
\ref{sec:coulomb} we analyze the Coulomb branch and its relation
to multi-center solutions; we also consider the connection with 
the AdS/CFT correspondence using linear perturbation theory. Next,
in Section \ref{sec:sugra}, we discuss the extreme rotating solutions 
to supergravity, and their underlying brane distribution. Section
\ref{sec:thermo} discusses the thermodynamics of rotating near-extreme
D3-brane black holes.  Following that, in Section \ref{sec:sing} we 
discuss the connection between singularities in spacetime and 
renormalisation group flows in the boundary theory. 
We conclude the paper by mentioning some connections 
with other recent work. The nonextreme rotating $D3$-brane solution is given 
in the Appendix.

\section{The Coulomb Branch and Multi-Center Solutions}
\label{sec:coulomb}

In this section we will establish a one-to-one correspondence between
points on the Coulomb branch of ${\cal N}=4$ super Yang-Mills theory and
multi-center D3-brane solutions of supergravity.   We begin by reviewing
some standard facts about the moduli space of the gauge theory, in particular
 its parametrization by  eigenvalues of Higgs fields or  by gauge invariant
operators.  An analogous discussion  of multi-center supergravity solutions 
follows. Next, we consider the relation between these ideas and the 
AdS/CFT correspondence relating the asymptotic behavior of bulk fields
to gauge theory operators; this discussion will exhibit an important 
subtlety.
Finally, we discuss a supergravity solution which includes a region of 
flat spacetime contained within a spherical shell of D3-branes.

\subsection{The Coulomb Branch}


 ${\cal N}=4$ $SU(N)$ 
super Yang-Mills theory includes the scalar fields 
$\Phi_i$ $(i=1 \ldots 6)$ which transform in
the vector representation of the $SO(6)$ R symmetry group, and
the adjoint representation of $SU(N)$.    The Coulomb branch corresponds
to giving these fields expectation values subject to the flatness
conditions $\left[ \Phi_i, \Phi_j \right] =0$.  Upon diagonalizing the
fields, the moduli space is parametrized by the $6N$ eigenvalues 
$y^{(a)}_i$     $(a=1 \ldots N)$:     
\be
\Phi_i = \left( \begin{array}{ccc} y^{(1)}_i & \mbox{} & \mbox{} \\
\mbox{} & \ddots & \mbox{} \\
\mbox{} & \mbox{} & y^{(N)}_i \end{array} \right) ~.
\label{higgs}
\ee
Tracelessness of $\Phi_i$ reduces the number of independent eignenvalues
to $6(N-1)$.
At generic points the gauge symmetry is broken to $U(1)^{N-1}$.  The
low energy effective theory valid below the scale set by the VEVs 
is obtained by integrating out massive off-diagonal degrees of freedom. 

For later comparison with the supergravity description we note that when $N$
is large, instead of giving a list of $6N$ eigenvalues, 
 it can
be more convenient to give an approximate description in terms of a 
continuous distribution.  Thus we let
$\sigma(\yv)$ denote the density of eigenvalues in the six dimensional
$\yv$ plane.

An alternative way to parametrize 
the moduli space is in terms of gauge invariant
operators. A complete set is given by the symmetric, traceless polynomials:
\be
{\cal O}_{(i_1 \cdots i_p)}= {\rm Tr}~ \Phi_{(i_1} \cdots \Phi_{i_p)}
\quad\quad\quad p= 2, \ldots ,N.
\ee
Operators ${\cal O}_{(i_1 \cdots i_p)}$ for $p>N$ can be expressed in
terms of operators with $p \le N$.  The descriptions in terms of eigenvalues
and in terms of gauge invariant operators contain the same 
information; given the eigenvalues one can work out the values of the
polynomials, and vice versa. For purposes of comparing with the supergravity
solutions below we also note that the Coulomb branch preserves ${\cal N}=4$
supersymmetry and the gauge coupling is not renormalised along this branch.

\subsection{Multi-Center D3-brane Solutions}

Configurations of the form:
\bea
ds^2 &=& H^{-{1\over 2}}_{D3}(-dt^2 + dx^2_1 + dx^2_2 + dx^2_3)
+H^{1\over 2}_{D3} \sum^6_{i=1}dy^2_i~, 
\label{eq:d3sol1}\\
C^{(4)} &=& (H^{-1}_{D3}-1)~dt\wedge dx_1 \wedge dx_2\wedge dx_3~,
\label{eq:d3sol2}
\eea
are solutions to type IIB supergravity for {\it any} harmonic function 
$H_{D3}(\vec{y})$. A general harmonic function has the integral representation: 
\be
H_{D3} ( \vec{y} ) = 1 + R^4~\int  
d^6 y^\prime {\sigma_{D3}(\vec{y^\prime})\over 
|\vec{y}-\vec{y^\prime}|^4}~,
\label{eq:hd3cart}
\ee
where $R^4$ is the constant:
\be
R^4 = 4\pi g_s \alpha^{\prime 2} N_{D3}~.
\ee
The distribution function $\sigma_{D3}$ 
is normalized as:
\be
\int d^6 y^\prime  ~\sigma_{D3}(\vec{y^\prime}) = 1~,
\label{eq:norm}
\ee
so $N_{D3}$ is the total number of D3-branes.

It is apparent that the space of multi-center solutions corresponds
to the possible distribution functions one can write down. The distribution:
\be
\sigma_{D3}(\vec{y}) = \frac{1}{N}\sum_{a=1}^N 
\delta^{(6)}(\vec{y}-\vec{y}^{(a)})~,
\ee
corresponds to D3-branes located at the discrete points $\vec{y}^{(a)}$.  
These points are naturally identified with the eigenvalues appearing
in (\ref{higgs}).  Thus one expects that the gauge theory and supergravity
configurations are dual descriptions.

In the classical supergravity limit the distribution need
not be a sum of $\delta$-functions, it may be continuous.   Indeed, the
solutions we obtain later from taking the extremal limit of certain
nonextremal solutions will be of this sort.  An important distinction
is that while discrete distributions always give rise to nonsingular solutions 
of classical supergravity, the solutions obtained from continuous 
distributions may exhibit naked singularities, as we discuss in 
Section 5. As mentioned previously, 
continuous distributions correspond in the gauge theory at finite $N$ 
to specifying the approximate distribution of eigenvalues.

Instead of directly giving the distribution $\sigma_{D3}(\yv)$ we can 
specify its moments: 
\be
{\cal O}_{(i_1 \cdots i_p)} = \int \! d^6y\, y_{(i_1} \cdots y_{i_p)}
\sigma_{D3}(\vec{y}).
\label{mom}
\ee
In terms of the moments, the harmonic function $H_{D3}$ 
takes the form of an expansion in spherical harmonics.  
Spherical harmonics on $S^5$ can be expressed in terms of symmetric, traceless
polynomials of $\yv$, so we write:
\be
H_{D3}(\yv) = 1+ R^4 \sum_{p=0}^{\infty} \frac{{\cal O}_{(i_1 \cdots i_p)}
Y_{(i_1 \cdots i_p)}(\Omega)}{|\yv|^{4+p}},
\label{eq:harmexpand}
\ee
where $\Omega$ denotes  angular coordinates on $S^5$.  
We note that in $SU(N)$ gauge theory
the first moment ${\cal O}_{i}$ actually vanishes, but we will not
indicate that explicitly.
The parametrization 
in terms of moments is analogous to the parametrization of the Coulomb branch
by gauge invariant operators. 

We see that the multi-center solutions, as described by the distribution
$\sigma_{D3}$ or its moments,  are in one-to-one correspondence with points
on the
Coulomb branch of the gauge theory, as parametrized by eigenvalues or by 
gauge invariant operators.
 In addition, the solutions (\ref{eq:d3sol1})-(\ref{eq:d3sol2})
preserve ${\cal N}=4$ supersymmetry with a constant dilaton.
 These facts strongly suggest that the Coulomb branch of the gauge theory is 
equivalent to the space of multi-center solutions
in the supergravity theory.
In the following we will see that agreement in the linearized
regime give further evidence for the correspondence.

\subsection{Supergravity Perturbations in $AdS_5\times S^5$}

The precise connection between gravity and gauge theory employs
the near-horizon limit of the supergravity solutions; {\it i.e.} 
the $1$ is omitted from the definition of the harmonic function $H_{D3}$, 
so that the resulting solutions are asymptotically $AdS_5 \times S^5$. 
The gauge theory operators are related to the deviations from 
$AdS_5 \times S^5$ as encoded in the asymptotic behavior of the supergravity 
fields~\cite{polyakovads,wittenads}. In the following we apply this formalism
to the multi-center solutions in order to determine the 
expectation values of  operators in the gauge theory. 
The goal is to verify the interpretation of multi-center solutions 
in the gauge theory as specifying points on the Coulomb branch of the 
moduli space, and in particular the precise map given in the discussion above.

The linearized supergravity perturbations on $AdS_5\times S^5$ were
classified in \cite{KRvN,gunmarcus} 
and their identification with gauge invariant
operators of ${\cal N}=4$ super Yang-Mills was made 
in~\cite{wittenads,fer1,fer2}. We focus on the fields 
corresponding to perturbations of the ten
dimensional metric, as they are the most relevant for our purposes.  
To make contact with \cite{KRvN} one writes the metric as:
\be
g_{{\hat\mu}{\hat\nu}} = \hat{g}_{{\hat\mu}{\hat\nu}} +
h_{{\hat\mu}{\hat \nu}},
\ee
where $\hat{g}_{{\hat\mu}{\hat\nu}}$ is the $AdS_5 \times S^5$ metric and
$h_{{\hat\mu}{\hat\nu}}$ is the perturbation.
To classify the perturbations, we divide the hatted indices into
$\mu$, denoting coordinates
in $AdS_5$, and $\alpha$, the coordinates on $S^5$. Thus the 
various perturbations are written as 
$h \equiv \hat{g}^{\alpha \beta}h_{\alpha \beta}$, 
$h_{(\alpha\beta)}$, $h_{\mu\alpha}$,
$h'_{\mu\nu} \equiv h_{\mu\nu}
+\frac{1}{3} \hat{g}_{\mu\nu} h$, where $(\alpha\beta)$ denotes a 
symmetric, traceless combination of indices.  
The next step is to expand the perturbations in $S^5$ harmonics and 
to insert the expansions into the quadratic order supergravity action,
to obtain a set of decoupled linear wave equations in $AdS_5$.    
Each independent partial wave component 
corresponds to a specific gauge theory operator.  
Thus by representing the multi-center solutions as perturbations around 
the background metric, it seems possible to identify which operators in 
the gauge theory have nonzero values. However, while this procedure 
works straightforwardly for the leading nontrivial harmonics, for the 
higher harmonics one needs to confront the fact that the multi-center 
solutions are actually solutions of {\em nonlinear} supergravity.

Let us make this more concrete.  From the form of a multi-center
solution, it is clear that the only perturbations which are potentially
nonvanishing are $h$ and $h'_{\mu\nu}$.  Let us first focus on $h$.
Expanding $h$ in harmonics on $S^5$ (where we now use a single k 
as shorthand for the indices $i_1 \ldots i_k$):
\be
h = \sum_k h_{k}(x^\mu) Y_k(\Omega),
\ee
one finds that to linear order the k'th harmonic obeys~\cite{KRvN}:
\be
[\Box_{AdS_5} - k(k-4) ]h_{k} =0.
\label{wave}
\ee
The general, normalizable solution  that is translationally invariant
in the brane direction $x_0 \ldots x_3$ is:
\be
h = \sum_k \frac{\tilde{Q}_k}{|\vec{y}|^{k}}Y_k(\Omega).
\label{eq:pert}
\ee
As discussed in \cite{wittenads}, the k'th harmonics of $h$ are dual in the
gauge theory to the k'th order symmetric, traceless polynomials
$O_{(i_1 \cdots i_k)}$.  In other words, as in~\cite{bkl}, the 
presence of the perturbation (\ref{eq:pert}) means that these gauge 
theory operators have expectation values:
\be
   <{\cal O}_{(i_1 \cdots i_k)}> \ \sim  \ \tilde{Q}_k.
\label{Oexp}
\ee
Thus one is studying the gauge theory at a point on the Coulomb branch.  
To add sources in the action for the operator $O_{(i_1 \cdots i_k)}$ 
one should in addition turn on a non-normalizable $h$ perturbation.

We now try to apply this approach to the multi-center solutions; we'll 
see that the analysis is more subtle than one might have guessed.
Consider a solution of the form (\ref{eq:d3sol1}) where the harmonic
function is:
\be
H_{D3} = \sum_{k=0}^{\infty} \frac{Q_k}{|\vec{y}|^{4+k}}Y_k(\Omega).
\ee
Keeping only the $Q_0$ term gives $AdS_5 \times S^5$ for the metric; the
perturbation $h$ is obtained by subtracting this contribution:
\begin{eqnarray}
h = \hat{g}^{\alpha \beta}h_{\alpha \beta} & = &
\frac{|\vec{y}|^2}{Q_0^{1/2}}\left\{\left( \sum_{k=0}^{\infty} 
\frac{Q_k}{|\vec{y}|^{4+k}}Y_k(\Omega)\right)^{1/2}
- 
\left(\frac{Q_0}{|\vec{y}|^{4}}Y_0(\Omega) \right)^{1/2}\right\} \\
&=& \frac{1}{2|\vec{y}|} \frac{Q_1}{Q_0}Y_1(\Omega)
+\frac{1}{2|\vec{y}|^2}\left(\frac{Q_2}{Q_0}Y_2(\Omega)
-\frac{1}{4} \frac{Q_1^{\, 2}}{Q_0^{\, 2}}\left(Y_1(\Omega)\right)^2\right)
 + \cdots  .
\label{hexpand}
\end{eqnarray}
There is a mismatch between this expression and the form given
in (\ref{eq:pert}): the angular dependence of the terms are
not spherical harmonics beyond the first term. Thus, beyond the 
lowest order the harmonics of $h$ do not obey the homogeneous equation 
(\ref{wave}), but rather some inhomogeneous equation:
\be
[\Box_{AdS_5} - k(k-4) ]h_{k} \approx  (h)^2 + (h)^3 +  \ \cdots,
\ee
where combinations of $k'<k$ harmonics appear on the right hand 
side (as do higher orders in the perturbations of other fields,
including the five-form). The reason for this complication is clear: 
in the linearized approximation of~\cite{KRvN} one --- by definition --- 
omits the source terms. However, one must verify that this is 
consistent, and this requirement translates into conditions on the underlying 
distribution function $\sigma_{D3}$. For a generic source of spatial extent
$\sim l$ we expect moments of the order $Q_k/Q_0 \sim l^k$ and then
the linear order perturbation theory
does {\it not} suffice. However, the linear approximation 
is valid for specially prepared distributions, and it always 
applies to the lowest harmonic contributing to the expansion 
of $h$ in (\ref{hexpand}).

In fact, in $SU(N)$ gauge theory the linear harmonic actually vanishes, 
$Q_1=0$. Thus the leading nontrivial order is $Y_2$. In this case the next
correction $Y_3$ cannot receive any corrections, because the terms
of lower order are of even degree. However, the $Y_4$ can receive 
non-linear corrections, of order $Y_2^2$, and the discussion proceeds 
as before.

In cases for which the linearized approximation is valid, one can readily
apply the methods of~\cite{bklt} to prove that the multi-center solutions
are described in gauge theory by points on the Coulomb branch.  The
idea is to consider a non-normalizable mode perturbation of $h_k$ around
a given solution.  On the supergravity side, upon integration by 
parts the change in action can be written as a boundary term proportional
to the normal derivative of $h_k$.  The variation of the gauge theory 
partition function is proportional to the expectation value of the
operator ${\cal O}_{(i_1 \cdots i_k)}$.  Equating these, one proves
(\ref{Oexp}).

So far we have only considered the field $h$.
However, similar considerations apply to other fields, in particular the 
perturbation $h'_{\mu\nu}$, which is dual to operators in the gauge theory 
involving products of Higgs fields with the energy-momentum tensor.  From the 
above analysis, one can reliably study the lowest harmonic of 
$h'_{\mu\nu}$ and thus read off the value of the energy-momentum tensor.  
In fact, one should actually compute the field $\phi_{(\mu\nu)}$, 
defined in (2.44) of~\cite{KRvN}, which differs from $h'_{\mu\nu}$ by 
terms related to $h$ and the five-form field. When these
effects are taken into account, direct evaluation in the case of a 
multi-center solution reveals that $\phi_{(\mu\nu)}$ vanishes, 
indicating zero energy-momentum density in the gauge theory. This is 
consistent with the interpretation of being on the Coulomb branch which 
--- being a supersymmetric configuration --- has zero energy.  

Thus the analysis of the multi-centered solutions in terms of perturbations
of $AdS_5 \times S^5$ is in accord with the Coulomb branch interpretation.  
As we have seen, though, trying to extract the actual values of all gauge 
invariant operators in the general case is not possible within the 
approximation of linearized supergravity.  However, this limitation
does not invalidate  our general prescription --- that the values of 
${\cal O}_{(i_1 \cdots i_k)}$ are equal to the moments of the distribution 
$\sigma_{D3}(\vec{y})$.

\subsection{Spherical Shell Solution}

\label{shell}
A multi-center solution of particular interest is obtained by taking the
D3-branes to be uniformly distributed over a five dimensional sphere
 of radius $|\vec{y}|=l$.  As is familiar from electrostatics, we expect
that inside the sphere the
 harmonic function will be constant, while outside the sphere it will
take the form corresponding to a D3-brane situated at the origin. 
Indeed by evaluating (\ref{eq:hd3cart}) (we drop the $1$ as appropriate
for comparing with gauge theory) we find:
\be
H_{D3}(\vec{y})  \ = \ \left\{ \begin{array}{ll}
R^4/l^4 &  \ |\vec{y}| < l \\
R^4/|\vec{y}|^4 & \ |\vec{y}| > l
\end{array} \right. 
\ee
  From the metric (\ref{eq:d3sol2}), we see that the solution consists of
flat ten dimensional Minkowski space inside the shell, and $AdS_5 \times
S^5$ outside.  Similarly, $C^{(4)}$ has vanishing field strength inside the
sphere, and its standard $AdS_5 \times S^5$ value outside.  We also note
that the divergence in the second derivative of $H_{D3}$ at $|\vec{y}|=l$
can  be smoothed out by instead considering a spherical shell of finite
thickness.  Although the  free parameter $l$ appears in the solution, 
note that its value can be rescaled by a coordinate transformation.
Indeed, the proper
radial size of the flat space region is $R$, and so is independent of $l$
\footnote{We thank O. Aharony for a discussion on this point.}.

This supergravity configuration is represented in the gauge theory as a 
particular point on the Coulomb branch.  
It is interesting to ask which gauge theory excitations represent the
fluctuations of supergravity fields in the flat spacetime region.  
We have not analyzed this question in detail. However, general features 
of the AdS/CFT correspondence suggest that bulk physics in the region 
near the origin is related to the low energy physics of the gauge theory.  
Thus we expect Minkowski space to control some nontrivial infrared fixed 
point. This interpretation merits further investigation.

\section{Supergravity Backgrounds}
\label{sec:sugra}
When we consider the gauge theory at finite temperature and chemical 
potential, the vacuum
degeneracy is removed. To determine the points in moduli space that 
are thus singled out, we consider rotating non-extreme backgrounds.
In the extreme limit, these solutions are non-rotating but the 
rotational parameters are retained. These are the nonvanishing moduli, 
and they allow a simple geometric interpretation. The rotating 
D3-brane is the main example, given in detail. 
For future reference, we also note that many extremal solutions have naked
singularities on surfaces containing D3- branes. The rotating M2- and 
M5-branes are qualitatively similar, and the results are stated with 
less discussion.

\subsection{Extreme Rotating D3-brane}

We are interested in configurations that arise as extreme limits of
rotating black 3-branes. The rotation group of the six-dimensional 
transverse space is the rank three group 
$SO(6)$, and so there are three independent rotational parameters, denoted 
$l_i$, $i=1,2,3$. Before taking the extreme limit these parameters
indicate angular momentum in three orthogonal two-planes. The five angular 
coordinates are chosen as the azimuthal angles $\phi_{1,2,3}$ of these
two-planes, and two additional polar angles $\theta$, $\psi$.
The full non-extreme solution is known explicitly and
it is reproduced in the Appendix. In the extreme limit
$m\rightarrow 0$ and $\delta\rightarrow\infty$, with 
$R^4\sim me^{2\delta}$ fixed, it becomes:
\bea
\label{eq:metric}
ds^2_S &=& H^{-{1\over 2}}_{D3} 
\left[ -dt^2 + dx^2_1 + dx^2_2 + dx^2_3\right]
+ H^{1\over 2}_{D3} f^{-1}_{D3} {dr^2\over \prod_{i=1}^3(1+{l^2_i\over 
r^2})}+\\ 
&+& H^{1\over 2}_{D3}r^2 \left[ ( 1 + {l^2_1 \cos^2\theta\over r^2}
+  {l^2_2 \sin^2\theta\sin^2\psi\over r^2}
+  {l^2_3 \sin^2\theta\cos^2\psi\over r^2})d\theta^2
+ \cos^2\theta d\psi^2 - \right. \nonumber \\
&-&
2 {l^2_2-l^2_3\over r^2}\cos\theta\sin\theta\cos\psi\sin\psi d\theta d\psi+ 
\nonumber \\
&+& \left. (1+{l^2_1\over r^2})\sin^2\theta d\phi_1^2
+ (1+{l^2_2\over r^2})\cos^2\theta\sin^2\psi d\phi_2^2
+ (1+{l^2_3\over r^2})\cos^2\theta\cos^2\psi d\phi_3^2 \right]~,\nonumber \\
C^{(4)} &=& (H_{D3}^{-1}-1)~dt \wedge dx_1 \wedge dx_2 \wedge dx_3~,
\eea
where: 
\bea
H_{D3} &=& 1 + f_D {R^4\over r^4}~, 
\label{eq:hd3rad}\\
f_{D3}^{-1}&=& ({\sin^2\theta\over 1 + {l^2_1\over r^2 }}+
{\cos^2\theta\sin^2\psi\over  1 + {l^2_2\over r^2 }}+
{\cos^2\theta\cos^2\psi\over 1 + {l^2_3\over r^2 }})\prod_{i=1}^3
( 1 + {l^2_i\over r^2} )~.
\eea
The main change relative to the non-extreme case is that 
the metric components of the form $g_{t\phi_i}$ now vanish. This
decouples the space within the brane from the space transverse to 
the brane, and it implies that the angular momenta of the solution
vanish. Despite these simplifications, the solution remains quite involved.
However, the configuration is extreme, and so it must be possible to 
write it in the general multi-center form given by 
(\ref{eq:d3sol1})-(\ref{eq:d3sol2}). 
Indeed, upon performing the coordinate change:
\bea
y_1 &=& \sqrt{r^2 + l^2_1} \sin\theta\cos\phi_1 \nonumber \\
y_2 &=& \sqrt{r^2 + l^2_1} \sin\theta\sin\phi_1 \nonumber\\
y_3 &=& \sqrt{r^2 + l^2_2} \cos\theta\sin\psi\cos\phi_2 \nonumber\\
y_4 &=& \sqrt{r^2 + l^2_2} \cos\theta\sin\psi\sin\phi_2 \nonumber\\
y_5 &=& \sqrt{r^2 + l^2_3} \cos\theta\cos\psi\cos\phi_3 \nonumber\\
y_6 &=& \sqrt{r^2 + l^2_3} \cos\theta\cos\psi\sin\phi_3~. 
\label{eq:ycoosD3}
\eea
the solution takes the form (\ref{eq:d3sol1})-(\ref{eq:d3sol2}) with the 
harmonic function $H_{D3}$ given by (\ref{eq:hd3rad}).
 It is straightforward to verify that $H_{D3}$ 
is indeed a harmonic function.  The fact that the coordinates 
(\ref{eq:ycoosD3}) greatly simplify the solution was noticed by 
Russo \cite{russorot} in the case $l_{2,3}=0$.

At the present stage the harmonic function (\ref{eq:hd3rad}) is still
given as a function of the Schwarzchild-like coordinates. Ideally,
we would like to rewrite it in terms of the isotropic coordinates
$\vec{y}$, and then identify the brane distribution underlying the 
extreme rotating solution by comparison with the general form 
(\ref{eq:hd3cart}). However, it does not in general appear practical 
to carry out these steps explicitly. Moreover, it follows 
from the coordinate change (\ref{eq:ycoosD3}) that the Schwarzschild-like 
coordinates only cover a subset of spacetime and so there may even
be an obstacle in principle: we do not {\it a priori} know the 
completion of the solution. 

In view of these difficulties we begin in the following with a simple 
special case where the underlying physics is simple, and then we
proceed towards increasing generality. In each step we first state the
result for the distribution function $\sigma_{D3}(\yv)$, 
and then discuss its justification.

\paragraph{One component:} Assume $l_2 = l_3 = 0$ but take $l_1$ arbitrary.
In this case the density of branes vanishes outside a disc of radius
$l_1$ in the plane defined by $y_3=y_4=y_5=y_6=0$. Moreover, the
distribution is uniform on the disc:
\be
\sigma_{D3}(\yv) = \frac{1}{\pi l_1^{\, 2}}\,\Theta(l_1 - 
\sqrt{y_1^{\, 2} +y_2^{\, 2}}~)~ \delta^{(4)}(\vec{y}_{\perp}).
\label{onell}
\ee

A first indication of this result follows by inspecting the form of $H_{D3}$ 
(\ref{eq:hd3rad}). In the present special case:
\be
r^4 f_{D3}^{-1} = r^2 ( r^2 + l^2_1 \cos^2\theta )~,
\ee
and so it is apparent that $H_{D3}$ has at least a
quadratic singularity when $r=0$, for all values of $\theta$. 
According to (\ref{eq:ycoosD3}) this translates into the surface  
$y_3=y_4=y_5=y_6=0$ (because $r=0$) and $y^2_1+y^2_2\leq l^2_1$ 
(because $\theta$ is arbitrary). Moreover, a two-dimensional 
surface charge in six spatial dimensions does indeed
give rise to quadratic singularities in the potential.

The result can be justified in detail by using
(\ref{eq:hd3cart}) to find the value of the harmonic function on any 
axis perpendicular to the plane of the disc:
\be
H_{D3} (\theta = 0 ) = 1 + R^4 \int_0^{l_1}
{1\over (\vec{y}^2+y^{\prime 2})^2} 
{2y^\prime dy^\prime\over l_1^2}
 = 1 + {R^4\over {\vec y}^2 ({\vec y}^2+l^2_1)}~.
\label{eq:hd3exp}
\ee
It is a simple matter to check that this equation
agrees with the general form (\ref{eq:hd3rad}) when the angular 
momenta satisfy $l_2=l_3=0$ and the angle $\theta=0$. It now follows 
from harmonic analysis that the two expressions 
agree also when $\theta\neq 0$. 
More explicitly, the harmonic property and the value of the function
at $\theta=0$ together determines the harmonic function everywhere as:
\bea
H_{D3} &=& 1 + \sum_{n=0}^\infty 
{Q_{2n}\over |{\vec y}|^{4+2n}}Y_{2n}(\cos^2\theta)~, \\
Q_{2n} &=& (-)^n R^4 l^{2n}_1~,
\eea
where the $Y_{k}(\cos^2\theta)$ are the scalar spherical harmonics on $S^5$,
with the normalization convention $Y_{k}(1)=1$.

The one component case considered here is the focus of
Section \ref{sec:thermo}. It is also the special case considered
by Russo~\cite{russorot}.

\paragraph{Two components:} Next, consider the case where $l_3$ = 0, but
$l_1$ and $l_2$ are arbitrary.   The branes are located at $y_5=y_6=0$, 
and on a three dimensional ellipsoidal surface:
\be
\sigma_{D3}(\vec{y})=\frac{1}{\pi^2 l_1^{\,2}l_2^{\, 2}} 
\delta({y^2_1+y^2_2\over l^2_1}+{y^2_3+y^2_4\over l^2_2}-1)
\delta^{(2)}(\vec{y}_{\perp})~.
\label{eq:el}
\ee

This result can be motivated as before, by considering the singularities
of the harmonic function. The full justification of the result also
proceeds as before.  The charged three dimensional 
surface gives the potential:
\bea
H_{D3}(\theta = \psi = 0 ) &=& 1 + R^4 \int^{\pi\over 2}_0 
{2\sin\theta^\prime \cos\theta^\prime  d\theta^\prime \over 
(l_1^2 \sin^2\theta^\prime  + l^2_2\cos^2\theta^\prime  + \vec{y}^2)^2}
\nonumber \\
&=& 1 + {R^4\over (\vec{y}^2+l^2_1)(\vec{y}^2+l^2_2)}~,
\eea
on any axis in the plane spanned by $y_5$ and $y_6$.
This expression agrees with the general form (\ref{eq:hd3rad}) when 
$l_3=0$ and $\theta=\psi=0$; and then harmonic analysis guarantees that 
the potential is correct throughout, as before.

The three dimensional surface defined by (\ref{eq:el}) is a 
closed surface when embedded in the four dimensional space with
$y_5=y_6=0$. In fact, it is immediately apparent that it is a generic
ellipsoid in four spatial dimensions. However, it is important
to note that, in the full six dimensional space, this three dimensional
surface does {\it not} divide space into two regions. In this sense
it is a higher dimensional analogue of a {\it ring}, because a ring 
divides the two dimensional plane into disconnected regions, leaving 
the three dimensional space connected (but not simply connected).

In the limit $l_2\rightarrow 0$ we should recover the one component
case considered above. According to (\ref{eq:el}) this limit forces 
$y_{3,4}\rightarrow 0$, but $(y^2_3+y^2_4)/l^2_2$ may remain finite.
To see the precise agreement it is easiest to integrate (\ref{onell}) and 
(\ref{eq:el}) with respect to $y_3, y_4$ and observe that the results 
match.

\paragraph{Three components (symmetric case):} 
An interesting special case with three nonvanishing components is
the symmetric assignment $l_1=l_2=l_3\equiv l$. Here the harmonic
function is simply:
\be
H_{D3} = 1 + {R^4\over (r^2 + l^2)^2} = 1 + {R^4 \over |\vec{y}|^4 }~.
\label{eq:hd3l3}
\ee
Thus the harmonic function is {\it independent} of the rotational parameter,
when written in terms of the isotropic coordinate $\vec{y}$.

However, this simple result does not determine the brane distribution
unambiguously: the original Schwarzschild-like coordinates only cover the 
part of spacetime with $r>0$; and so the equation above can be 
applied only when $|\vec{y}|>l$. Thus, the underlying brane distribution
can be any spherically symmetric distribution with the correct
total charge. In particular, the potential in (\ref{eq:hd3l3}) could
arise from a point source, or alternatively from a {\it sphere} of 
radius $l$. The latter interpretation corresponds to the solution discussed
in Section \ref{shell}.

\paragraph{Three components (general case):} 
when all three components of the angular momentum are nonvanishing 
the harmonic function does not exhibit any singularity in the limit 
of $r\rightarrow 0$. This behavior is compatible with a brane distribution 
that is some five dimensional surface, perhaps with nonvanishing density 
in its interior. The ellipsoid with the defining equation:
\be
{y^2_1+y^2_2\over l^2_1}+{y^2_3+y^2_4\over l^2_2}+
{y^2_5+y^2_6\over l^2_3}=1~,
\label{eq:el2}
\ee
is the surface $r=0$. It realizes many symmetries of the problem and
we suspect that it plays some preferred role. However, as noted above in 
the special case $l_1=l_2=l_3$, the underlying brane distribution is an 
ambiguous entity. The reason is that the Schwarzschild-like coordinates only 
apply outside the surface (\ref{eq:el2}), and we cannot extend the solution 
inside this surface without {\it assuming} a particular distribution 
of the charges.

\subsection{Extreme Rotating M5-brane}
A distribution of M5-branes is described by the solution:
\bea
ds^2 &=& H_{M5}^{-{1\over 3}}
(-dt^2 + dx^2_1 + dx^2_2 + dx^2_3+ dx^2_4 + dx^2_5)
+H_{M5}^{2\over 3} \sum_{i=1}^5 dy^2_i~,
\label{eq:m5g} \\
C^{\star(3)} &=& (H_{M5}^{-1}-1) ~
dt\wedge dx_1 \wedge dx_2\wedge dx_3\wedge dx_4
\wedge dx_5~.
\label{eq:m5c}
\eea
In general, the harmonic function $H_{M5}$ is given in terms of the 
normalized distribution of M5-branes ($\sigma_{M5}$) as:
\be
H_{M5} ( \vec{y} ) = 1 + R^3~\int  
d^5 y^\prime {\sigma_{M5}(\vec{y^\prime})\over 
|\vec{y}-\vec{y^\prime}|^3}~,
\ee
where the coefficient of the harmonic function is $R^3={1\over 2}l_p^3 N_{M5}$
\footnote{We use units where the eleven dimensional Planck length is given by 
$l_p = (2\pi g_s)^{1\over 3}\sqrt{\alpha^\prime}$. However, the precise
numerical factors will play no role.}.

The space transverse to the M5 is five dimensional. Its angular space is 
therefore classified by the rank two group $SO(5)$, giving 
two independent angular momenta, parametrized by $l_1$ and $l_2$. The
azimuthal angles in the planes of these angular momenta are $\phi_{1,2}$,
and the remaining angular coordinates are the polar angles $\theta$
and $\psi$. The extreme rotating M5 is given in~\cite{cy96c}
using Schwarzchild-like coordinates. In the isotropic coordinate
system defined by: 
\bea
y_1 &=& \sqrt{r^2 + l^2_1} \sin\theta\cos\phi_1 \nonumber \\
y_2 &=& \sqrt{r^2 + l^2_1} \sin\theta\sin\phi_1 \nonumber\\
y_3 &=& \sqrt{r^2 + l^2_2} \cos\theta\sin\psi\cos\phi_2 \nonumber\\
y_4 &=& \sqrt{r^2 + l^2_2} \cos\theta\sin\psi\sin\phi_2 \nonumber\\
y_5 &=& r\cos\theta\cos\psi~,
\label{eq:ycoosM5}
\eea
it is identical to 
the canonical solution (\ref{eq:m5g}-\ref{eq:m5c}) with
the harmonic function:
\be
H_{M5} = 1 + f_{M5}~{R^3\over r^3}~,
\label{eq:hm5rad}
\ee
where:
\be
f_{M5}^{-1} = ({\sin^2\theta\over {1+{l^2_1\over r^2}}}+
{\cos^2\theta\sin^2\psi\over {1+{l^2_2\over r^2}}}+\cos^2\theta \cos^2\psi)
(1+{l^2_1\over r^2})(1+{l^2_2\over r^2})~.
\ee
It is not manifest that (\ref{eq:hm5rad}) is a harmonic function,
because it is written in terms of Schwarzchild-like coordinates,
but this is nevertheless the case.

Next, we identify the underlying brane distribution, assuming that 
one component of the angular momentum vanishes, say $l_2=0$.
The singularities of the harmonic function shows that the branes are
confined to the disc defined by $y_3=y_4=y_5=0$ and 
$|\vec{y}_\parallel |\equiv \sqrt{y^2_1+y^2_2} \leq l_1 $. A
simple computation verifies that the precise distribution is: 
\be
\sigma_{M5}(\vec{y}) = 
{1 \over 2\pi l_1\sqrt{l^2_1- |\vec{y}_\parallel|^2 }}~
\Theta(l_1-|\vec{y}_\parallel|)~\delta^{(3)}({\vec{y}_\perp})~,
\ee
where $\vec{y}_\perp$ denote the three components $y_{3,4,5}$. 
Note that this distribution is {\it not} uniform: 
the density of branes diverges at the boundary of the disc. 
Roughly, the branes form a ring of radius 
$l_1$, but not a sharp one: the density falls off smoothly
between the peak at the ``ring'' and the centre.

\subsection{Extreme Rotating M2-brane}
The solution that describes any collection of M2-branes is:
\bea
ds^2 &=& H_{M2}^{-{2\over 3}}(-dt^2 + dx^2_1 + dx^2_2)
+H_{M2}^{1\over 3} \sum^{8}_{i=1}dy^2_i~,
\label{eq:m2g}\\
C^{(3)} &=& (H^{-1}_{M2}-1)~dt\wedge dx_1 \wedge dx_2~,
\label{eq:m2c}
\eea
where $H_{M2}$ is a harmonic function that we write in the form:
\be
H_{M2} ( \vec{y} ) = 1 + R^6~\int  d^8 y^\prime 
{\sigma_{M2}(\vec{y^\prime})\over 
|\vec{y}-\vec{y^\prime}|^8}~.
\ee
The constant $R^6$ is related to the number of M2-branes as 
$R^6=8l_p^6 N_{M2}$. 

The space transverse to the M2-brane is 8-dimensional and so there are 
four independent angular momenta and seven angular coordinates,
denoted $\theta$,$\psi_{1,2}$, $\phi_{1,2,3,4}$. 
Starting from (\ref{eq:m2g}-\ref{eq:m2c}), the change of coordinates:
\bea
y_1 &=& \sqrt{r^2 + l^2_1} \sin\theta\cos\phi_1 \nonumber \\
y_2 &=& \sqrt{r^2 + l^2_1} \sin\theta\sin\phi_1 \nonumber\\
y_3 &=& \sqrt{r^2 + l^2_2} \cos\theta\sin\psi_1\cos\phi_2 \nonumber\\
y_4 &=& \sqrt{r^2 + l^2_2} \cos\theta\sin\psi_1\sin\phi_2 \nonumber\\
y_5 &=& \sqrt{r^2 + l^2_3} \cos\theta\cos\psi_1\sin\psi_2\cos\phi_3 \nonumber\\
y_6 &=& \sqrt{r^2 + l^2_3} \cos\theta\cos\psi_1\sin\psi_2\sin\phi_3 \nonumber\\
y_7 &=& \sqrt{r^2 + l^2_4} \cos\theta\cos\psi_1\cos\psi_2\cos\phi_4 \nonumber\\
y_8 &=& \sqrt{r^2 + l^2_4} \cos\theta\cos\psi_1\cos\psi_2\sin\phi_4
\label{eq:ycoosM2}
\eea
yields the form of the extreme rotating M2 brane given in~\cite{cy96c}.
This computation also identifies the relevant harmonic function as:
\be
H_{M2} = 1 + f_{M2}~{R^6\over r^6}
\ee
where 
\bea
f_{M2}^{-1} &=& G_{M2} \prod_{i=1}^4 (1+{l^2_i\over r^2}) \\
G_{M2} &=& {\sin^2\theta\over {1+{l^2_1\over r^2}}}+
{\cos^2\theta\sin^2\psi_1\over {1+{l^2_2\over r^2}}}
+{\cos^2\theta\cos^2\psi_1\sin^2\psi_2\over {1+{l^2_3\over r^2}}}
+{\cos^2\theta\cos^2\psi_1\cos^2\psi_2\over {1+{l^2_4\over r^2}}}
\nonumber
\eea
It is elementary (but tedious) to show that $H_2$ is indeed harmonic, 
{\it i.e.} it satisfies the Laplace equation.

We have determined the underlying distribution of branes in the
case of a single angular momentum parameter $l_1$, {\it i.e.} 
$l_2=l_3=l_4=0$. It is:
\be
\sigma_{M2}(\vec{y^\prime}) = 
{4  ( l^2_1 - |\vec{y}_\parallel |^2)\over 2\pi l^4_1}~
\Theta(l_1-|\vec{y}_\parallel|)~
\delta^{(6)}({\vec{y}_\perp})
\ee
where the two dimensional vector $\vec{y}_\parallel$ is within the plane 
of rotation. Thus, we find a disc with radius $l_1$ in the
$M2$ case too. The distribution is nonuniform, with no sharp peaks. 
In particular, it is smooth at the boundary of the disc.

\section{Thermodynamics of Rotating D3-branes}
\label{sec:thermo}

In this section we study the thermodynamics of spinning D3 branes
in some detail.  We also attempt to relate the results obtained from 
supergravity to the expected behavior of the dual 
Yang-Mills theory. Direct calculations in the strongly coupled
Yang-Mills theory are difficult, but we will see that 
in some respects the qualitative behavior  agrees with our expectations. 
 In other respects   the supergravity theory reveals non-trivial 
features about the gauge theory.  For example, in one limiting case 
there is evidence that the excitations in the gauge theory are governed by 
a string theory with a string tension determined by the scale of the Higgs 
VEVs. 

\subsection{The Supergravity Theory}
For the sake of simplicity, we focus throughout this section 
on the case where the D3 branes rotate in only one plane, {\it i.e.} 
when only one of the  three angular momentum parameters is  nonzero.  
The metric describing this brane configuration is the special case
$l_2=l_3=0$ of the metric discussed in the Appendix (\ref{eq:metric}). 
Various thermodynamic properties can be read off from the metric. 
The number of branes $N$ is related to the AdS curvature scale $R$ by:
\be
\label{nofbranes}
R^4 = ({2 \over \pi^4}  G_N)^{1 \over 2} N  = 2 m ~\cosh\delta~\sinh\delta. 
\ee
The mass  and angular momentum are:
\bea
\label{russomassam}
M&=&{\pi^2 \over 8} ~{L^3 m \over G_N}~ (4\cosh^2 \delta +1), \\
\label{russoangmom}
J&=&{\pi^2 \over 4}~ {L^3 \over G_N}~  lm\cosh \delta,
\eea
where $L$  is the size of  each direction along which the branes extend. 

The position of the horizon in the coordinates of (\ref{eq:metric}) 
is given by:
\be
\label{russohorizon}
r_H^2={\sqrt{l^4 +8m} -l^2 \over 2}.
\ee
The entropy, temperature and angular velocity of the horizon are then given 
by~\cite{gkp,cy96d}: 
\bea
\label{russoentropy}
S&=&{\pi^3 \over 2 } {L^3 \over G_N} m r_H  \cosh\delta \\ 
\label{russotemp}
T&=&{r_H \over 4 \pi m \cosh\delta} \sqrt{l^4 + 8m},\\
\label{russoangvel}
\Omega_H&=& {l r_H^2 \over 2m\cosh\delta}~.
\eea
One can verify that these  quantities satisfy the thermodynamic relation:
\be
\label{fl}
TdS= dE - \Omega_H dJ.
\ee

In the discussion below we consider the near-extremal limit  defined by taking
$\delta$ large with $N$ fixed. The thermodynamic properties in this limit 
follow from (\ref{nofbranes}--\ref{russoangvel}); they are:
\bea
\label{nenergy}
E&=&{3  \pi^2\over 8} {L^3 \over G_N} m~, \\
\label{nangmom}
J&=&{\pi^2 \over 8} {L^3 \over G_N} ~l \sqrt{2m}~R^2~,\\
\label{nentropy}
S&=&{\pi^3 \over 4}{L^3 \over G_N}~r_H \sqrt{2m}~R^2~,\\
\label{ntemp}
T&=&{1\over 2\pi} {r_H \sqrt{l^4+8m}\over \sqrt{2m}~R^2}~,\\
\label{nangvel}
\Omega_H&=& {l r_H^2 \over \sqrt{2m}~R^2}~,
\eea
where $r_H$ is still given by (\ref{russohorizon}).

For purposes of comparing with the dual Yang-Mills theory
we are interested in  black hole solutions  which asymptote to  AdS spacetime. 
These can be obtained from the black brane solutions asymptotic to 
flat space in two steps.  First, one takes the near extremal limit  where
$m,l$ are small in units of $R$. Next, following~\cite{juanads},  
one takes the  near horizon limit  of this near-extremal 
solution.  Once the black hole solutions
have been obtained in this manner though,    
the condition on $m$  and $ l$ being
small can be relaxed.   One finds that  
the resulting metric is a solution of the
Einstein equations asymptotic to AdS space,  
for all values of the parameters 
$m$ and $l$.  Moreover, the thermodynamic properties of  these solutions  
continue to be given by (\ref{nenergy})-(\ref{nangvel}) for
the entire range of parameters (with the energy and angular momentum being
defined with respect to asymptotic AdS space).

\subsection{The Yang-Mills Theory.}

We now turn to understanding these thermodynamic properties from the
Yang-Mills theory point of view. The angular momentum (\ref{nenergy})
corresponds to  charge under an $SO(2)$ subgroup of the $SO(6)$ R-symmetry 
group of the Yang-Mills theory. The thermodynamic quantities in 
(\ref{nenergy}) are in general complicated functions of the dimensionless 
parameter $l^4/m$. 
We will not be able to reproduce these  functions  by a direct
calculation in the strongly coupled gauge theory. 
Instead, we will study  two limiting cases,  when  $l^4/m$ tends to zero and
infinity,   and show
how  some qualitative features agree  with our expectations.
We will then comment on the general case towards the end.  

\subsubsection{The $l^4/m \ll 1 $ limit}

When $l=0$, we have a non-rotating black hole. It is well known that
in this case the dependence of the entropy on the temperature is accurately 
given by a calculation in the free field limit, apart from an overall 
normalisation~\cite{gkp}. This leads one to try to understand  the 
 $l^4/m \ll1$ case   in this limit  as well. 
Usually,  the Grand Canonical ensemble with fixed chemical potential,
rather than fixed charge, is more convenient  for studying  problems
of this kind.  So  one could  hope that  the thermodynamics  with a  small 
chemical potential  can be understood  by working in the  non-interacting 
theory. 

However, this is an inconsistent starting point:
a massless relativistic Bose gas cannot sustain a non-zero chemical 
potential; any attempt to turn on a chemical
potential gives rise to Bose Einstein condensation. Mathematically, 
a chemical potential gives rise to negative occupational probabilities for
some zero modes of the scalar fields.   

Despite this fact one can proceed in the following simple-minded manner.
We turn on a chemical potential in the non-interacting theory,
and neglect the zero-modes in studying the thermodynamics. 
An elementary computation gives the entropy:
\be
\label{entrgt}
S={2\pi^2\over 3}T^3 L^3 N^2~
\left(1+{3 \over 4 \pi^2}~{\Omega_H^2\over T^2}\right).
\ee
Here $\Omega_H$  denotes the chemical potential in the Yang Mills theory, 
which is identified with the angular velocity in the supergravity theory
(\ref{nangvel}). In contrast, expanding the supergravity formulas
(\ref{nenergy})-(\ref{nangvel}) one finds:  
\be 
\label{entrsugra} S={\pi^2 \over 2} T^3 L^3 N^2 \left(1+ {1 \over 2 \pi^2}
  {\Omega_H^2 \over T^2}  \right) 
\ee
The comparison between the $\Omega_H$ independent terms in (\ref{entrgt})
and (\ref{entrsugra}) is well known~\cite{gkp}. 
Here we see that the free field 
calculation correctly reproduces the functional 
dependence, in particular the factor $N^2$, for the subleading term as well. 
The disagreement in the numerical coefficients is not surprising because
(\ref{entrgt}) and (\ref{entrsugra}) correspond to Yang-Mills theory at
weak and strong 't Hooft coupling, respectively.

The naive treatment of the chemical potential described above may be 
better justified than one suspects at first glance. To see this we first 
note that the need to include interactions once the chemical potential is 
non-zero has a simple physical explanation from the point of view of D3 branes.
If the D3- branes are  non-interacting, any attempt to turn on an angular 
momentum or an angular velocity is unsustainable --- 
the branes simply fly apart in the absence of any forces between them.  
However, once interactions are
turned on and the D3 branes are excited above extremality,  they 
experience a gravitational attraction which could provide the required
centripetal force for rotation. 

These considerations suggest that, in the Yang-Mills theory, keeping the 
one loop interactions could be a useful starting point for the analysis. 
We have not carried out such an analysis. However, one expects that the 
resulting Higgs VEVs are small when the angular momentum is small 
and further that the rotational energy is a small fraction of the total 
energy of the system. This suggests that the effects of the Bose condensate 
can in fact be neglected when calculating in this limit and that the 
one-loop analysis should compare favorably with (\ref{entrgt}). 

\subsubsection{The $l^4/m \gg 1 $ limit}
\label{sec:large}

As the angular momentum is increased one expects the effects of the 
Bose condensate to grow. The D3 branes  should be typically displaced from 
the origin,  
and a reasonable fraction of their total energy should go into the Bose 
condensate, giving rise to the rotation, while the rest goes into thermal 
excitations, accounting for the entropy. 

First, we consider the strict limit, where $m \rightarrow 0$ with $l$ kept 
fixed. Here we have a very explicit description of the condensate:
it  corresponds to the configuration discussed in section (\ref{sec:sugra}), 
where the D3 branes are distributed uniformly on a planar disc.
One finds  from (\ref{nenergy})-(\ref{nangvel}) that the energy $E$ 
and angular momentum vanish in this limit, as expected.
The entropy and the angular velocity vanish as well, while 
the temperature is constant. Since the curvature diverges at the 
horizon in this limit, we should not attribute too much significance 
to the behaviour of the entropy, temperature and angular velocity. Even so,  
it is reassuring to note that the entropy approaches zero signalling 
that the system settles down into the ground state. 

Next, consider a slightly  non-extremal configuration, 
for which $m/l^4 \ll 1$.  
This should be described by a slightly excited version of the uniformly
charged disc. For any given angular momentum and total energy there is 
some  consistent distribution of energy between the rotational and thermal 
excitations that could be determined, in principle, by a variational  
calculation. The thermal excitations give rise to an attractive 
gravitational force that in turn  provides the centripetal force to
spin the D3 branes. 

There is a simple estimate that supports this picture. From (\ref{nenergy}) 
and (\ref{nangmom}) we find that the total energy is related to the 
angular momentum as:
\be
\label{exactenang}
E= {J^2\over 2I}~,
\ee
where $I$ is the ``effective moment of inertia'' given by:
\be
\label{effi}
I= {1 \over 6} M_0 l^2~,
\ee
and $M_0={1 \over 4} ({2 \over G_N})^{1/2} N L^3$ is the total mass of the $N$
D3 branes. 
(For future reference we note that (\ref{exactenang}) and 
(\ref{effi}) are valid for all values of $l^4/m$.) 
By way of comparison, the moment of inertia of a collection of branes,
uniformly distributed on a disk of radius $l$, is: 
\be
\label{idisk}
I={1 \over 2} M_0 l^2. 
\ee
This is of the same form as (\ref{idisk}), but it is numerically smaller
as is to be expected: first, the branes should be somewhat denser near 
the origin as compared to the edges; second, in (\ref{effi}) 
the energy $E$ is the total energy of the system instead of the rotational
energy, which should be some fraction of this. The details of this
picture should be given by a variational principle.

We should also mention that the moment of inertia is exactly 
of the form one would expect from a Bose-Einstein condensate in 
the gauge theory.  (\ref{effi}) can be rexpressed in terms of the typical 
vacuum expectation value for the Higgs fields $v \sim \alpha^{'} l$ as
$I \sim  N v^2/ g_s$. In the gauge theory we expect a state 
which carries charge to correspond to a time dependent vacuum expectation 
value for the Higgs field of the form $\phi \sim v e^{-i \omega t}$. 
 After accounting for a factor of $N$ due to a color trace this gives rise
to an energy  $E \sim N v^2 \omega^2/ g_s$ --- exactly 
in accord with (\ref{exactenang})--(\ref{effi}). 

As for the entropy, in the limit $m\ll l^4$ it is  given by:
\be
\label{limentro}
S \sim (g_sN)^{1/2}  E/v ,
\ee
where $v$, as above, is the scale of the VEVs for the two non-zero Higgs 
fields in the gauge theory. The linear dependence of the entropy on the 
energy is suggestive. It indicates that the 
effective theory  for  low-energy excitations above the Bose condensate is 
a string theory  with a  string tension of order $v^2/(g_s N)$. 
Understanding this string theory is  obviously  of interest, but it might be
challenging since it clearly involves the strongly coupled nature of the 
gauge theory. In Section \ref{sec:sing}
 we will find that the energy regime governed by the 
effective string theory corresponds in the bulk to a singularity in the 
metric. 

\subsubsection{Concluding Remarks}

We end this section with some remarks. 
As was noted above (\ref{exactenang})--(\ref{effi}) are in fact valid  
everywhere in parameter space. This suggests that $l$ continues to provide 
a measure of the size of the brane configuration, even away from the
limit $l^4 \gg m$. In particular $l$ decreases as the energy increases 
for a fixed value of the angular momentum, so the brane configuration 
shrinks towards the origin of moduli space. When the angular momentum 
approaches zero the branes move to the origin and we recover the 
description valid when $l^4\ll m$. 

Throughout this section we have considered the Yang-Mills theory living on a
three  torus and we note that in this case the thermodynamic formulas 
(\ref{nenergy})-(\ref{nangvel})  do not indicate a phase transition 
in the  bulk of the parameter space governed by
$m$ and $l$ \footnote{The phase transition discussed in~\cite{gubserrot}
occurs as a function of the temperature and chemical potential at a point 
which lies on the boundary of the  parameter space spanned by $l,m$.}.  
In contrast, for Yang-Mills theory on a  three sphere a phase 
transition does occur, in the microcanonical ensemble,  once the radius of the 
curvature $r_H \sim R$. This is the direct analogue of the phase transition 
discussed in~\cite{bdhm} for the non-rotating case and is associated 
with the dominant configuration changing from a black hole in AdS space to  
one which is localised in ten dimensions. Such a phase transition does not 
occur in the toroidal case~\cite{peetross},  
in accord with one's expectations 
from no-hair theorems.
This is easy to see for the rotating case as well. The relevant 
comparison here is between two seven dimensional black holes,
one smeared and the other localised in the $S^3$ transverse to the plane of 
rotation. 

\section{Gauge Theory Interpretation of the Spacetime Singularity}
\label{sec:sing}

In some cases the extreme solutions discussed in Section \ref{sec:sugra} are 
singular on surfaces containing D3 branes. In this section we relate 
the appearance of this  singularity to  a change in the behavior of the 
gauge theory under renormalization group flow.
For simplicity we focus on the extreme D3 brane solution with one non-zero
angular momentum parameter, {\it i.e.} $l_1 \ne 0, l_{2,3}=0$. 
The discussion employs the Schwarzchild-like coordinates of 
(\ref{eq:metricextreme}).

The extreme brane geometry has a singularity at radial coordinate $r=0$.
For small $r$ the curvature invariant $R_{\mu \nu} R^{\mu \nu}$ diverges like 
$l^2/(R^4 r^2)$\cite{russorot}. Thus the curvature scale is of order 
the string scale at radial position $r\sim l/(gN)^{1/2}$.

We will need to relate scales in the bulk and the boundary theories;
there are two relations of this kind~\cite{peetpol}. 
The scale of the VEVs for the Higgs fields is related to the
size of the brane configuration by $v= l/\alpha^\prime$. A second 
relation connects the cutoff in the gauge theory to a radial size \cite{wisu}. 
In AdS space this takes the form:
\be
\label{iruv}
L= R^2/r,
\ee
where $r$ and $L$  are the radial position in the bulk and length
scale on the boundary respectively.
This relation is also  the appropriate one when supergravity 
modes are used as probes~\cite{bklt, peetpol}, or when the string world sheet 
{\it ansatz} is used to evaluate Wilson loops in the bulk theory 
\cite{maldaw, rey}. In the latter case (\ref{iruv}) relates
the size of the loop on the boundary to the minimum radial position  
to which it meanders inwards in the bulk.

The brane geometry under discusion here is not AdS. However,
when $r_{min}\gg l$ one can still use the AdS geometry in estimating 
the dynamical scale --- thus the ${\cal N}=4$ UV theory governs the behavior 
of the gauge theory. At $r_{min} \sim l$ the full geometry comes into 
play; this corresponds to a scale $L \sim R^2/l$ in the gauge theory. 
At this point the qualitative behavior of a probing string world sheet 
changes: a rough estimate shows that a further decrease in $r_{min}$ 
does not lead to a substantial increase in the size of the loop. 
As a result when $r_{min} = l/(gN)^{1/2}$ --- the radial position
at which the curvature is of string scale ---  the loop is still of
order $L \sim R^2/l$.

Let us pause to note that in terms of energies on the boundary the 
approximately AdS geometry continues to suffice until an energy scale
$E \sim {l \over \sqrt{gN} \alpha^\prime}$. This is lower than the scale  
of the VEVs for some of the Higgs fields $v$. Thus we learn, from 
the supergravity calculation, that in this regime the Wilson loop 
continues to behave as in the superconformal ${\cal N}=4$ theory;
in particular the energy between two
static color sources scales like $E \sim 1/L$.

We now turn to our main interest, the spacetime singularity.
It is clear that from the above arguments that from the gauge theory 
point of view the singularity is related to physics at the energy 
scale $E_{sing}={l \over \sqrt{gN} \alpha^\prime}$. 
In Section  \ref{sec:large} we studied the low-energy excitations of the gauge 
theory using thermodynamics and presented evidence that
it is governed by an effective string theory. The string tension of this
effective theory is set by $E_{sing}$. An interesting picture therefore 
emerges. The high energy behavior of the gauge theory  is characteristic
of the ultraviolet behavior of the superconformal ${\cal N}=4$ theory. Its 
low-energy behavior is instead governed by an effective string theory. 
The cross-over between the two should occur at a scale of order the 
string tension, which agrees with the radial location of the singularity 
in the bulk.

Clasical supergravity is clearly inadequate for going
``past'' the singularity. In contrast, the gauge theory is clearly valid at
lower energies as well. In fact at the scale 
$E \sim l/(\sqrt{N}\alpha^\prime)$
all the nonabelian gauge bosons become massive and the theory
reduces to weakly coupled $U(1)^{N-1}$ gauge theory.

\section{Discussion}
We conclude the paper with a few comments on other recent work.

Rotating D3-brane solutions have been analyzed in the computation 
of glueball masses in~\cite{russorot,russoetal}. Taking the angular 
momentum parameters to be large allows one to decouple certain unwanted 
states from the theory. According to our results this limit amounts
to considering a collection of D3-branes distributed on a large disk 
of radius $l$. This corresponds to a specific symmetry breaking
pattern of the $SU(N)$ gauge theory, still leaving a nontrivial 
theory. Its precise nature will be relevant for the interpretation of the 
result given in~\cite{russorot,russoetal}.

Some aspects of the AdS/CFT correspondence on the Coulomb branch were
studied in~\cite{bulkbranes}. In particular, one can go to a point 
in moduli space with unbroken $SU(N-2) \times U(1) \times U(1)$ gauge 
symmetry and compute the effective action of the $U(1)$ gauge fields.  
The result can be interpreted as the interaction between two D3-branes in
$AdS_5 \times S^5$ due to the exchange of supergravity quanta.  Linearized
supergravity is adequate here since for large $N$ the presence of the two
D3-branes in the bulk is a small perturbation of the background.  By
contrast, in our examples we consider geometries which are
large deformations of $AdS_5 \times S^5$ in the bulk.  

Finally we mention~\cite{gubserrot}, on the thermodynamics of rotating
D3-branes. This work has some overlap with section~\ref{sec:thermo},
which was essentially completed when~\cite{gubserrot} appeared.

\vspace{0.2in} {\bf Acknowledgments:} We thank Sumit Das, Jerome
Gauntlett, and Emil Martinec for helpful comments. 
We also thank O. Aharony, J. Maldacena and R. Myers for 
comments on the first version of this paper.
PK and FL are supported in part by DOE Grant no. DE FG02 90ER 40560.
PK is also supported by the NSF Grant no. PHY 9600697 and
FL is also supported by a McCormick fellowship through the EFI. 
ST is supported in part by DOE grant DE-AC02-76CH0300.

\appendix
\section{The nonextreme rotating D3-brane}

The black hole solutions with one or several rotational parameters are
quite complicated in general. In~\cite{horsen,cy96d} a large class
of solutions were found that correspond to rotating fundamental strings 
in arbitrary dimension $D$. After duality transformations the solutions 
with $D=7$ can be brought into a form where the only excited $U(1)$ 
field is the one coupling to the charge of $D3$-branes. 
Taking into account
the various scalar fields that are also present, the general rotating
D3-brane solution in 10 dimensions can be computed. The
resulting metric is:
\bea
ds^2_S &=& H^{-{1\over 2}}_{D3} \left[ - 
H_{D3} dt^2 + dx^2_1 + dx^2_2 + dx^2_3+ \right.  
\label{eq:metricextreme} \\
&+& \left. f_D {2m\over r^4} (\cosh\delta dt-l_1 \sin^2\theta d\phi_1
-l_2 \cos^2\theta \sin^2\psi d\phi_2-l_3 \cos^2\theta \cos^2\psi d\phi_2)^2
\right] \nonumber \\
&+&  H^{1\over 2}_{D3} f^{-1}_{D3} {dr^2\over \prod_{i=1}^3(1+{l^2_i\over r^2})
-{2m\over r^4}}+  \nonumber \\ 
&+& H^{1\over 2}_{D3}r^2 \left[ ( 1 + {l^2_1 \cos^2\theta\over r^2}
+  {l^2_2 \sin^2\theta\sin^2\psi\over r^2}
+  {l^2_3 \sin^2\theta\sin^2\psi\over r^2})d\theta^2
+ \cos^2\theta d\psi^2 + \right. \nonumber \\
&-&
2 {l^2_2-l^2_3\over r^2}\cos\theta\sin\theta\cos\psi\sin\psi d\theta d\psi+ 
\nonumber \\
&+& \left. (1+{l^2_1\over r^2})\sin^2\theta d\phi_1^2
+ (1+{l^2_2\over r^2})\cos^2\theta\sin^2\psi d\phi_2^2
+ (1+{l^2_3\over r^2})\cos^2\theta\cos^2\psi d\phi_3^2 \right]~,
\nonumber 
\eea
where: 
\bea
H_{D3} &=& 1 + f_D {2m\sinh^2\delta\over r^4} \\
f_{D3}^{-1}&=& ({\sin^2\theta\over 1 + {l^2_1\over r^2 }}+
{\cos^2\theta\sin^2\psi\over  1 + {l^2_2\over r^2 }}+
{\cos^2\theta\cos^2\psi\over 1 + {l^2_3\over r^2 }})\prod_{i=1}^3
( 1 + {l^2_i\over r^2} )~.
\eea
The only matter field that is excited is the four-form gauge field: 
\bea
C^{(4)} &=&
- (H_{D3}^{-1}-1)~{1\over\sinh\delta}~dx_1 \wedge dx_2 \wedge dx_3\wedge
\\
&\wedge& (\cosh\delta dt-l_1 \sin^2\theta d\phi_1
-l_2 \cos^2\theta \sin^2\psi d\phi_2-l_3 \cos^2\theta \cos^2\psi 
d\phi_2)~.
\nonumber
\eea
In particular, the dilaton field in ten dimensions is constant.
The total D3-brane charge is:
\be
R^4 = 2m\sinh\delta\cosh\delta = 4\pi g_s \alpha^{\prime 2} N_{D3}~.
\ee

The mass and angular momentum can be read off from the asymptotic
geometry. This gives the formulae (\ref{russomassam}--\ref{russoangmom}).



\end{document}